\documentclass[preprint2,10pt]{aastex}
\usepackage{textcomp,amsmath,txfonts}
\usepackage{natbib}
\begin{document}

\sloppy

\renewcommand*{\phi}{\varphi}
\renewcommand*{\epsilon}{\varepsilon}
\newcommand{\total}{\operatorname{d}\!}
\renewcommand{\arraystretch}{1.10}
\renewcommand{\vec}[1]{\mathbf{#1}}
\tabcolsep 3pt

\newcommand{\mum}{\,\hbox{\textmu{}m}}
\newcommand{\mm}{\,\hbox{mm}}
\newcommand{\cm}{\,\hbox{cm}}
\newcommand{\m}{\,\hbox{m}}
\newcommand{\km}{\,\hbox{km}}
\newcommand{\AU}{\,\hbox{AU}}
\newcommand{\pc}{\,\hbox{pc}}
\newcommand{\Kelvin}{\,\hbox{K}}
\newcommand{\second}{\,\hbox{s}}
\newcommand{\yr}{\,\hbox{yr}}
\newcommand{\Myr}{\,\hbox{Myr}}
\newcommand{\Gyr}{\,\hbox{Gyr}}
\newcommand{\rad}{\,\hbox{rad}}
\newcommand{\mJy}{\,\hbox{mJy}}
\newcommand{\magn}{\,\hbox{mag}}
\newcommand{\scmperccm}{\,\hbox{cm}^{2}\hbox{cm}^{-3}}
\newcommand{\persmpers}{\,\hbox{m}^{-2}\hbox{s}^{-1}}
\newcommand{\g}{\,\hbox{g}}
\newcommand{\gperccm}{\,\hbox{g\,cm}^{-3}}
\newcommand{\ergperg}{\,\hbox{erg\,g}^{-1}}
\newcommand{\gpers}{\,\hbox{gs}^{-1}}
\newcommand{\kmpers}{\,\hbox{km\,s}^{-1}}

\title{Will New Horizons see dust clumps in the Edgeworth-Kuiper belt?}

\author{Christian Vitense}
\and
\author{Alexander V. Krivov}
\and
\author{Torsten L\"ohne}
\affil{Astrophysikalisches Institut, Friedrich-Schiller-Universit\"at Jena,
           Schillerg\"a{\ss}chen~ 2--3, 07745 Jena, Germany}
\email{vitense@astro.uni-jena.de}

\begin{abstract}
Debris disks are thought to be sculptured by neighboring planets.
The same is true for the Edgeworth-Kuiper debris disk, yet no direct observational 
evidence for signatures of giant planets in the Kuiper belt dust distribution has been 
found so far.
Here we model the dust distribution in the outer solar system
to reproduce the dust impact rates onto the dust detector onboard the New 
Horizons spacecraft measured so far and to predict the rates during the Neptune 
orbit traverse.
To this end, we take a realistic distribution of transneptunian 
objects to launch a sufficient 
number of dust grains of different sizes and follow their orbits by
including radiation pressure, Poynting-Robertson and stellar wind drag,
as well as the perturbations of four giant planets.
In a subsequent statistical analysis, we calculate number densities and lifetimes of
the dust grains in order to simulate a collisional cascade.
In contrast to the previous work, our model not only considers
collisional elimination of particles, but also includes production of finer debris.
We find that particles captured in the $3$:$2$ resonance with Neptune build 
clumps that are not removed by collisions, because the depleting effect of collisions is counteracted by production of 
smaller fragments.
Our model successfully reproduces the dust impact rates measured by New Horizons 
out to $\approx{}23\AU$ and predicts an increase of the impact rate of about a factor of 
two or three around the Neptune orbit crossing.
This result is robust with respect to the variation of the vaguely known 
number of dust-producing scattered disk objects,
collisional outcomes, and the dust properties.
\end{abstract}


\shortauthors{Vitense et al.}
\shorttitle{The Kuiper Belt dust and New Horizons}

\maketitle

\section{INTRODUCTION}

The symbiosis between debris disks and planets is multifaceted.
Not only do both represent natural outcomes of the planetesimal and planet accretion processes,
it has also long been realized that planets should sculpt the disks by their gravity,
and thus the 
observed structure in debris disks can be used as a tracer of planets. 
In one case~--- $\beta$~Pic~--- the observed disk structure
has been undoubtedly attributed to interactions with a directly imaged planet 
\citep{Lagrange-et-al-2009,Lagrange-et-al-2010,Lagrange-et-al-2012a}.
In a few other cases, such as Fomalhaut \citep{Kalas-et-al-2008,Kalas-et-al-2013}, HR~8799 
\citep{Su-et-al-2009}, and HD 95086 \citep{Rameau-et-al-2013a}, the relation between the 
disks and directly imaged planets has yet to be understood.
Many more systems have been found where resolved debris disks reveal
various types of structure, possibly driven by gravity of unseen planets orbiting in the disks' 
inner cavities, exemplified by the clumpy disk of $\epsilon$~Eri
\citep{Greaves-et-al-2005} or spirals in the HD~141569 disk \citep{Wyatt-2005b}.
However, some of these features are yet to be 
confirmed observationally, for instance requiring disambiguation with possible background
objects. Also, models of planet-induced structure in the disk suffer from many 
uncertainties, especially those arising from poorly known dust properties and from difficulties of 
including collisions into the models. Besides, alternative explanations
of the observed disk structure that do not require the presence of planets are possible 
\citep[e.g.][]{Artymowicz-Clampin-1997,Grigorieva-et-al-2007,Debes-et-al-2009}.

It is natural and tempting to look at our solar system.
Here, the orbits and masses of planets are precisely known.
Largely known are also the populations of parent bodies, maintaining
its debris disk, which include asteroids, comets, and Edgeworth-Kuiper belt (EKB) objects (EKBOs).
Less well known are the properties of the dust cloud
that these parent bodies replenish and maintain.
Yet a dust ring encompassing the Earth orbit has been discovered and its structure 
successfully ascribed to resonant interactions between dust grains and the Earth 
\citep{Dermott-et-al-1994}. In the outer solar system, it has been predicted that Neptune should
create two dusty clumps just exterior to its orbits, one ahead and one behind
the planet slightly outside its orbit \citep{Liou-Zook-1999}. The simulated images of 
those clumps
from the \citet{Liou-Zook-1999} paper have been used in dozens of subsequent papers as a vivid 
prototype of what can also be expected in extrasolar debris disks.

However, the predicted EKB clumps are still lacking direct observational confirmation.
Remote observations of thermal emission are hampered
by an extremely low dust density in the EKB debris disk
\citep{Vitense-et-al-2012}, which causes the
foreground emission of the zodiacal cloud to outshine the thermal flux from 
the EKB dust.
Luckily, the New Horizons spacecraft is now on its way to the outer solar system.
Onboard, beside several other instruments, is the Venetia Burney Student Dust Counter (SDC),
capable of measuring impacts of grains between $10^{-12}\g< m < 10^{-9}\g$ 
in mass \citep{Horanyi-et-al-2008}.
New Horizons will traverse the trailing clump around early 2015 and might see a dust 
enhancement there.

This paper addresses the question of whether the SDC has chances to detect
the dust clump, proving its existence, and whether the expected data can be 
used to constrain the models.
So far, there have been several attempts to model the EKB dust clumps
\citep[e.g.,][]{Liou-Zook-1999,MoroMartin-Malhotra-2002,MoroMartin-Malhotra-2003,Kuchner-Stark-2010,Poppe-et-al-2010,Han-et-al-2011}.
However, most of these studies ignored possible effects of grain-grain collisions.
\citet{Kuchner-Stark-2010} did include them with the aid of their novel ``collisional grooming''
algorithm. Nevertheless, the collisions in their model
only acted to eliminate the colliders; it is not surprizing therefore that the collisions in 
their model tends to erase the clumps. Including collisional {\em production} of fine debris
might counteract this, putting dust enhancements back into play; we check this in this 
paper.
Besides, none of the previous studies provided detailed predictions for the dust impact rates
for the SDC detector aboard New Horizons along its trajectory; we do this here.
We also try to use a more realistic ``true'' distribution of the EKB objects, acting as parent 
bodies for the dust \citep{Vitense-et-al-2010},
and to include perturbations of all four giant planets.

Section~\ref{sec:dust_production} describes the procedure and results of the 
modeling in a collisionless approximation. Section 3 does the same with the collisions included.
Conclusions are made in section 4 and discussed in section~5.

\section{COLLISIONLESS MODEL}\label{sec:dust_production}

\subsection{Method}

In the spirit of previous studies \citep[e.g.,][]{Liou-Zook-1999,MoroMartin-Malhotra-2003, 
Kuchner-Stark-2010}
we released dust grains from EKBOs and followed them with numerical integrations.
The first step was to set the distribution of the parent bodies.
Instead of using an artificial distribution, we invoked the debiased EKB distribution
of \citeauthor{Vitense-et-al-2010} (\citeyear{Vitense-et-al-2010}, their Figure~5). 
They developed an algorithm to remove observational biases
from the orbital elements of EKBOs
listed in various EKBO discovery surveys.
The corrected distributions of semimajor axis $a$, eccentricity $e$, and inclination $i$
in each of three populations as defined in their paper
(classical Kuiper belt, CKB; resonant objects, RES; and scattered objects, SDO)
were approximated by gaussian functions.
The mean values $\overline{a}$, $\overline{e}$, and $\overline{\imath}$ along with
the standard deviations $\sigma_a$, $\sigma_e$, and $\sigma_i$
are listed
in Table~\ref{tab:parent_body_distribution}.
The longitude of the ascending node, the argument of pericenter, and the mean 
anomaly were assumed to be uniformly distributed between $0^\circ$ and $360^\circ$. 

\begin{table}[h!]
\caption{Initial orbital distribution for parent bodies for classical (CKB), resonant (RES) and scattered (SDO) objects.
}
\centering
\begin{tabular}{c c c c c c c}
\hline\hline
Population & $\overline{a}$ [AU] & $\sigma_a$ [AU] & $\overline{e}$ & $\sigma_e$ & $\overline{\imath}$ [$^\circ$] & $\sigma_i$ [$^\circ$] \\
\hline
CKB        & $43$ & $2.5$             & $0.0\,\,$ & $0.1\,\,$ & $\,\,0$ & $17$\\
RES        & $39$ & $0.5$             & $0.2\,\,$ & $0.1\,\,$ & $15$    & $10$\\
SDO        & $60$ & $10\quad\,$       & $0.45$    & $0.15$    & $20$    & $10$\\
\hline
\end{tabular}
\label{tab:parent_body_distribution}
\end{table}

From each of the three populations we started $1000$ single particles of nine different sizes, 
selected as described below, so that a total of $27000$ particles were handled.
After release, each particle feels direct solar 
radiation pressure, causing the initial grain orbit to differ from that of the parent body
\citep{Burns-et-al-1979}.
The particle orbit acquires a larger semimajor axis
and typically, but not always, a larger eccentricity than those of the parent EKBO.
We have included this effect with the aid of Eqs. (19)-(20) of \citet{Krivov-et-al-2006}.

The orbital evolution of particles was followed by numerically integrating their
equations of motion with a Bulirsch-Stoer routine with an adaptive stepsize control
and an accuracy parameter of $10^{-11}$ \citep{Press-et-al-1992}.
We took into account perturbations from
Neptune, Uranus, Saturn, and Jupiter (the planets did not interact mutually), 
the Poynting-Robertson (PR) force,
as well as the stellar wind drag which was assumed to have the strength of $33\%$ 
of the PR effect \citep{Gustafson-1994}.
The integration was stopped when the particle was ejected from the system ($r > 300\AU$)
or impacted onto a planet or the sun.
The position $(x,y,z)$, velocity $(v_x,v_y,v_z)$, and time $t$ since release 
of the dust grains were recorded once every orbit of Neptune,
as was also done in the previous studies
\citep{Liou-Zook-1999,MoroMartin-Malhotra-2003,Kuchner-Stark-2010}.
Since the dust production and loss rates are assumed to be constant, the system is ergodic.
Therefore, different records of the same simulated particle 
can be interpreted as the same-time records of different physical grains launched at
different time instants in the past.
This effectively increases the number of records of our simulations to 
$\approx 2\times 10^9$ for all three parent populations and nine sizes.

To set the sizes of the particles for simulations, we made use of the ratio of 
direct radiation pressure and gravity, $\beta$  \citep{Burns-et-al-1979}.
The radiation pressure efficiency for the grains
was calculated using the Bruggeman mixing rule and standard Mie theory 
\citep{Bohren-Huffman-1983}, assuming three different 
mixtures of astrosilicate \citep{Laor-Draine-1993} and water ice
\cite{Warren-1984}.
These were a 50\%-50\% astrosilicate-ice mixture with a 
bulk density of $\rho = 2.35\gperccm$,
a 10\%-90\% astrosilicate-ice mixture ($\rho = 1.43\gperccm$),
and pure ice ($\rho = 1.00\gperccm$).
We then selected nine equally spaced logarithmic size (or mass) bins
covering the mass range which can be detected by the dust counter onboard the 
New Horizons spacecraft \citep[$10^{-12}\g < m < 10^{-9}\g$,][]{Horanyi-et-al-2008}.
The list of $\beta$ values and corresponding grain sizes and masses
for the three mixtures is given in Table~\ref{tab:used_betas}.

\begin{table}[h!]
\caption{$\beta$-ratios, sizes, and masses of grains used in the simulations
for different material mixtures.
}
\centering
\tabcolsep 2pt
\begin{tabular}{c c c c c c c c}
\hline\hline
\#   & $\beta$        & \multicolumn{2}{c}{$\rho = 2.35\gperccm$}& \multicolumn{2}{c}{$\rho = 1.43\gperccm$}& \multicolumn{2}{c}{$\rho = 1.00\gperccm$}\\
 &  & $s$ [\textmu{}m] & $m$ [g] & $s$ [\textmu{}m] & $m$ [g] & $s$ [\textmu{}m] & $m$ [g]\\
\hline
$0$ & $0.576$ & $0.43$     & $7.5\mathrm{E}{-13}$       & $0.30$  &  $1.7\mathrm{E}{-13}$     & $0.46$  & $4.1\mathrm{E}{-13}$\\
$1$ & $0.404$ & $0.65$     & $2.7\mathrm{E}{-12}$       & $0.65$  &  $1.6\mathrm{E}{-12}$     & $0.69$  & $1.4\mathrm{E}{-12}$\\
$2$ & $0.259$ & $0.99$     & $9.5\mathrm{E}{-12}$       & $1.05$  &  $7.0\mathrm{E}{-12}$     & $1.25$  & $8.2\mathrm{E}{-12}$\\
$3$ & $0.164$ & $1.51\,\,$ & $3.4\mathrm{E}{-11}$       & $1.60$  &  $2.5\mathrm{E}{-11}$     & $1.78$  & $2.4\mathrm{E}{-11}$\\
$4$ & $0.106$ & $2.30\,\,$ & $1.2\mathrm{E}{-10}$       & $2.45$  &  $8.8\mathrm{E}{-11}$     & $2.57$  & $7.1\mathrm{E}{-11}$\\
$5$ & $0.070$ & $3.51\,\,$ & $4.3\mathrm{E}{-10}$       & $3.60$  &  $2.8\mathrm{E}{-10}$     & $3.45$  & $1.7\mathrm{E}{-10}$\\
$6$ & $0.046$ & $5.42\,\,$ & $1.6\mathrm{E}{-9}\,$      & $5.80$  &  $1.2\mathrm{E}{-9}$      & $4.60$  & $4.1\mathrm{E}{-10}$\\
$7$ & $0.029$ & $8.59\,\,$ & $6.2\mathrm{E}{-9}\,$      & $10.3$  &  $6.6\mathrm{E}{-9}$      & $6.76$  & $1.3\mathrm{E}{-9}$\\
$8$ & $0.018$ & $14.3$     & $2.9\mathrm{E}{-8}\,$      & $18.3$  &  $3.8\mathrm{E}{-8}$      & $10.0$  & $4.2\mathrm{E}{-9}$\\
\hline
\end{tabular}
\label{tab:used_betas}
\end{table}

\subsection{Results}

\begin{figure*}
  \begin{center}
  \includegraphics[width=1.0\textwidth]{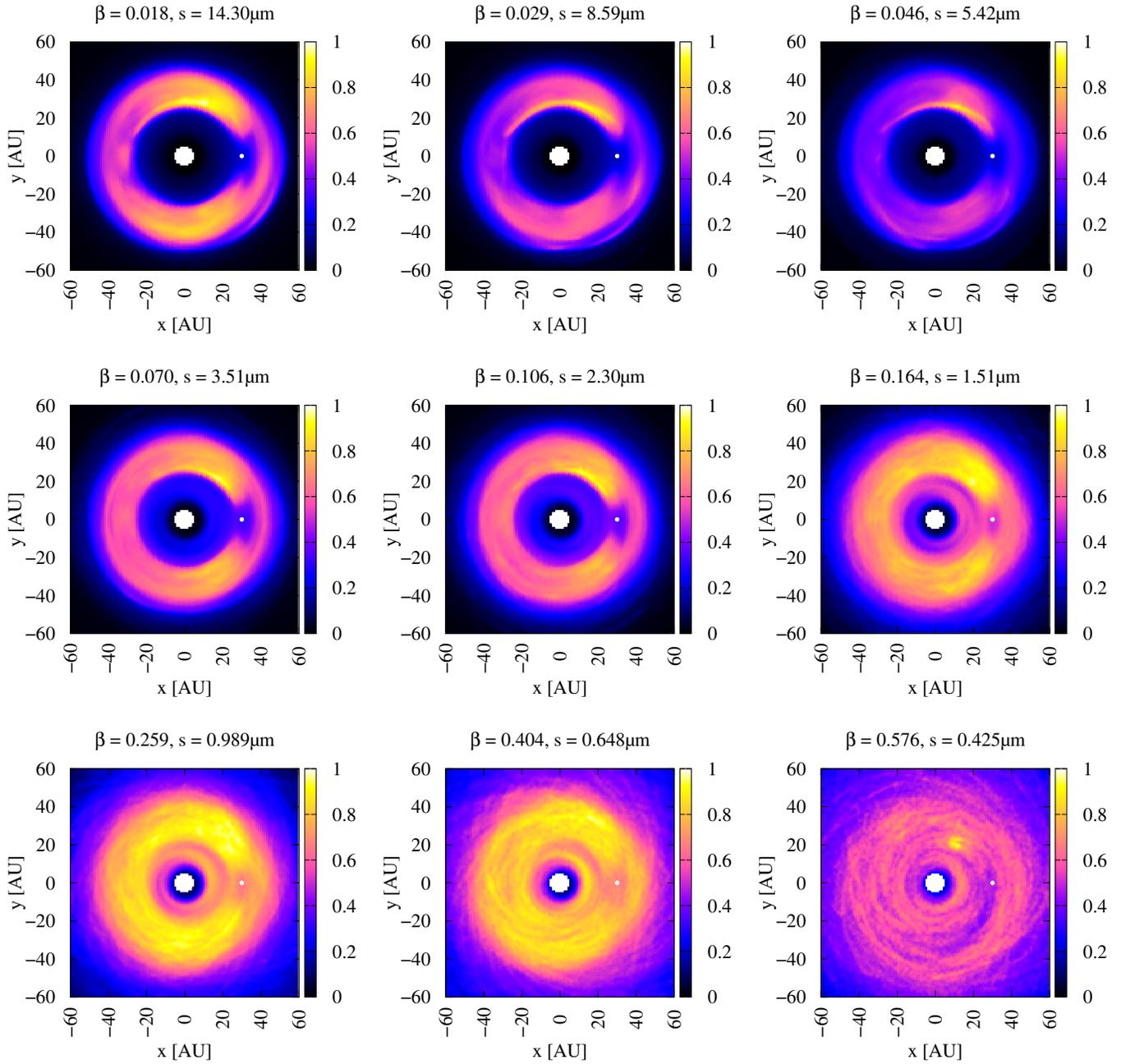}\\
  \end{center}
  \caption{
    Number density in the ecliptic plane normalized to the densest bin (each 
    size separately) for 
    collisionless particles of different sizes.
    The small white dot is the location of Neptune.
    }
  \label{fig:map_no_coll}
\end{figure*}

We binned the positions of particles, separately for each of the sizes,
into an $(x,y)$-grid with $\Delta x = \Delta y = 1\AU$.
The numbers of records in each bin originating from CKB, RES, and SDO
were weighted as 1 : 0.2 : 1, which is approximately the mass ratio of these populations
in the ``true'' EKB of \citet{Vitense-et-al-2010}.
The resulting snapshots of the dust distribution are shown in
Figure~\ref{fig:map_no_coll}.
These are nearly identical to those published and discussed in detail previously
\citep[e.g., by][]{Liou-Zook-1999,Kuchner-Stark-2010}, and thus are summarized below only 
briefly.

Most of the dust is confined to a ring-like structure
between slightly interior to the Neptune orbit and the outer edge of the classical EKB.
This ring has an extended gap around the position of Neptune, and two broad clumps
created by grains trapped in mean-motion resonances with Neptune.
These resonant structures are more pronounced for larger particles
since their probability to be captured into a mean-motion resonance is higher 
\citep[e.g.,][]{Wyatt-2003,Mustill-Wyatt-2011}.
The distribution of small particles, just above the blowout limit, is more uniform.
Note that not all of the $\beta=0.576$ particles in Fig.~\ref{fig:map_no_coll} are
in blowout orbits, because the parent body orbits are generally eccentric.
A grain released from an eccentric orbit near the aphelion can stay in bound orbit
even if $\beta > 0.5$.

Outside the main belt, the number density rapidly decreases,
reflecting the fact that even the dust grains produced by the scattered objects are located 
most of their time in the main belt.
Inside the main belt, the dust density also drops significantly.
Noteworthy, though not sufficiently discussed in the previous papers, is
the fact that the clumps of the $3:2$ resonance 
have relatively sharp inner edges slightly interior to the Neptune orbit,
at ($25$--$27)\AU$.
First, our analysis of a number of individual orbits has shown that the grains trapped 
in the resonance maintain the resonant value of $a$, while their orbital eccentricities 
$e$ grow, so that the pericentic distances $q$ decrease.
The growth in $e$ gradually slows down at some point, and so does the decrease in $q$.
As a result, many resonant grains maintain $q \approx (25$--$27)\AU$ over a long period 
of time. Second, a swarm of grains in Keplerian ellipses sharing the same $q$ is 
known to reach a large (formally infinitely large) number density at the location of 
their pericenters, because the radial velocity of grains $\dot{r}=0$ there
\citep[e.g., Eq. 2.13 in][]{Krivov-et-al-2005}.
Both effects together explain the 
sharp inner edges seen in Fig.~\ref{fig:map_no_coll}.
Jupiter, Saturn and, to lesser degree, Uranus contribute to the inner gap opening 
by gravitational scattering of dust drifting inward from the EKB.

\section{MODEL WITH COLLISIONS}\label{sec:collisions}

\subsection{Method}

\begin{enumerate}

\item 

As in Section~\ref{sec:dust_production}, we start
with distributing all the grains of size $s$ into $(x,y)$-bins.
This time, however, we will need to calculate the collisional rates and so
need to know the absolute number of different-sized particles
in the system.
Accordingly, we assume an initial size distribution $\propto s^{-q}$
and scale every particle by a certain normalizing factor $\xi$.
Every grain (i.e., every record) is counted
\begin{equation}\label{eq:xi}
 \xi\left(\frac{s}{s_\text{max}}\right)^{-q+1}
\end{equation}
times, where $s_\text{max}$ is the largest grain size we considered.
Although we will see later that the initial distribution of grains is
overwritten by the collisional fragments, we assumed $q = 3.5$
which is the exponent for an idealized disk in a collisional equilibrium \citep{Dohnanyi-1969}.
The $+1$ in the exponent is due to our using logarithmic size bins.
This is done for each size listed in Table~\ref{tab:used_betas}.

\item

Treating the collisions, we assume them to occur between the colliders of the 
same size, so that the following procedure is done for every grain size separately.
In our numerical integrations, we keep record of the velocity of every grain
to calculate the mean relative velocity $\overline{v_{\text{rel}}}$ 
between \emph{all} grains of size $s$ in every spatial bin.
Denoting by $n$ the total number of records in an $(x,y)$ bin obtained in step $1$,
the relative velocity is given by
\begin{equation}
 \overline{v_{\text{rel}}}^2   = \frac{1}{n^2}\sum_{i,j=1}^{n}\left(\vec{v}_{i} - \vec{v}_{j}\right)^2
                               = \frac{2}{n}\sum_{i=1}^{n} \vec{v}_{i}^2 
                               - \frac{2}{n^2}\left(\sum_{i=1}^{n} \vec{v}_{i}\right)^2,
\end{equation}
where the indices $i,j$ represent the single records of all particles%
\footnote{Strictly speaking, the summation over $i,j$ should run from 
$1$ to $n$ with $i\neq j$. Excluding $i=j$ would change the prefactor to
$[n(n-1)]^{-1}$, but $n$ is so large that the correction is negligible.}.
Typical relative velocities between two like-sized particles in the EKB 
region are $(3-4)\kmpers$ outside the clumps 
and $\lesssim 2\kmpers$ inside them.
Next, the mean collisional time in every $(x,y)$ bin is calculated as
\begin{equation}\label{eq:lifetime}
 t_{\text{life}}^{-1} = n\sigma \overline{v_{\text{rel}}},
\end{equation}
where $\sigma = 4\pi s^2$ is the collisional cross section for two colliders of radius $s$.
Step $2$ results in the knowledge of lifetime
for every size in every $(x,y)$-bin, and we can start including collisions.

\item

Once the lifetime is known, we repeated step $1$
but this time counted every particle $\exp{(-t/t_\text{life})}$ times, where
$t$ is the time elapsed after release from the parent body.
This takes into account the loss due to collisions.
Then we checked whether the grains of size $s$ in question
can be produced in collisions of larger grains $s_\text{larger}$
and if so, the number of particles which are produced by larger grains was \emph{added}.
This was implemented as follows.
When two particles of the same size collide with each other,
the mass of the largest collisional fragment is
\citep{Krivov-et-al-2006}
\begin{equation}\label{eq:m_x}
 m_x \approx m_\mathrm{target}\frac{Q^*_\mathrm{D}(m_\mathrm{target})}{\overline{v_\mathrm{rel}}^2}
\end{equation}
where $m_\text{target}$ is the mass of one of the two colliders and
$Q_\mathrm{D}^*$ is the critical specific energy which is needed to disrupt a particle%
\footnote{A particle is considered disrupted when the largest fragment
contains less than half of the mass of the original particle.} \citep[e.g.,][]{Krivov-et-al-2005}:
\begin{equation}
 Q_\text{D}^* = A (s/1\m)^{3b}.
\label{QD}
\end{equation}
As a nominal case, we took $A = 10^6\ergperg$, $3b = -0.37$ 
\citep[e.g.,][]{Benz-Asphaug-1999}, but also tried 
other values to test how strongly the results can be affected by a poorly known strength of 
the dust material. Note that our largest dust grains are still far from
being held together by gravity, so the gravity part of the equation is neglected
and we only consider the strength regime.
If the impact energy $E_\text{imp} = 0.25 m_\mathrm{target} v_\mathrm{rel}^2$ 
exceeds $2 m_\mathrm{target} Q_\mathrm{D}^*$ (two objects of the same size have to be 
destroyed), the total mass of the target is fragmented, and we can write
\begin{align}
 m_\mathrm{target}&=\int\frac{4}{3}\pi\rho s^3 \total N(s)
                   = \int_0^{s_x}\frac{4}{3}\pi\rho s^3 N_0 \left(\frac{s}{s_x}\right)^{-p}\total s\nonumber\\
                  &=\frac{4}{3}\pi\rho N_0\frac{s_x^4}{4-p}
\end{align}
with $s_x$ being the radius of the largest fragment (cf. Eq.~\ref{eq:m_x}).
Here, $\total N(s)$ is the fragment distribution function, $N_0$ its normalization 
constant, and the exponent $p$ is assumed to be $3.5$ \citep{Fujiwara-1986}.
With this fragment distribution it is easy to calculate
how many small grains of size $s$ are produced: 
\begin{equation}
  n_\text{prod} = \left(\frac{s_\text{larger}}{s}\right)^{3}\left[\left(\frac{s_+}{s_x}\right)^{4-p}-\left(\frac{s_-}{s_x}\right)^{4-p}\right].
\end{equation}
Since we use discrete sizes, $s_-$ and $s_+$ are the minimum and maximum physical sizes 
represented by the size bin around $s$.
We recall that $n_\text{prod}$ is the number of grains $s$ produced by 
\emph{one} larger grain $s_\text{larger}$. 
Hence, we have to multiply it with the total number of larger grains $n_{\text{larger}}$
in the same $(x,y)$-bin.
Thus the number of the newly produced particles is
\begin{equation} 
   n_\text{prod} n_{\text{larger}}[1-\exp(-t/t_{\mathrm{life,larger}})] ,
\end{equation}
where $[1-\exp(-t/t_{\mathrm{life,larger}})]$ is the fraction
of larger grains already disrupted by collisions at time $t$.
Finally, for one size in one $(x,y)$-bin we count every particle (i.e., every record)
\begin{equation} 
   \xi\left(\frac{s}{s_\text{max}}\right)^{-q}
   \underbrace{ \exp{(-t/t_\text{life})}}_\text{loss}
+  \underbrace{n_\text{prod} n_{\text{larger}}[1-\exp(-t/t_{\mathrm{life,larger}})]}_\text{gain}
\end{equation}
times. The gain term represents the particles which are produced by \emph{all} particles
larger than themselves.%
\footnote{Note that the ``gain'' term has to be divided by the number of records of like-sized particles
in a certain $(x,y)$-bin from the collisionless analysis to avoid double counting.
}
Note that this simple model tacitly assumes that small particles are produced by large
ones locally. In particular, it does not include the radiation pressure effect on 
the newly produced particles immediately upon release. Nor does it include the drift of 
small particles inward from their birth location.
This simplification is discussed in more detail in Sect.~\ref{sec:discussion}.

\end{enumerate}

Steps $2$ and $3$ should be executed repeatedly, until the changes from iteration to 
iteration in each $(x,y)$-bin become smaller than the required accuracy.
However, the convergence turned out to be so fast that already the first iteration
ensures the accuracy at a $<10$\% level. For this reason, in the calculations presented here 
we applied collisions only once.

Having calculated the number density, we can compute the
particle flux on the SDC. It is given by
\begin{equation}
 F = \int_{10^{-12}\mathrm{g}}^{10^{-9}\mathrm{g}} m \, \hat{n} \, v_\mathrm{imp}\total\ \ln m ,
\label{eq:flux} 
\end{equation}
where $m$ is the mass of the dust particle,
$\hat{n}$ is the number density per logarithmic mass $m$,
and $v_\mathrm{imp}$ is their average velocity relative to the spacecraft.
In practice, we replaced integration with summation over all mass bins
$[m_-, m_+]$ that overlap fully or partly with the SDC sensitivity range
$[10^{-12}\g, 10^{-9}\g]$.
A logarithmic interpolation was applied between the bins 
and to the minimum ($10^{-12}\g$)
and maximum ($10^{-9}\g$) detectable masses, since these
do not coincide exactly with the bin boundaries $m_-$ and $m_+$.
Furthermore, in Eq.~(\ref{eq:flux}) we replaced $v_\mathrm{imp}$
with the heliocentric velocity of New Horizons, $v_\mathrm{NH}$.
The latter was set to $v_\mathrm{NH} = 15.5\kmpers$, the velocity New Horizons had shortly 
after the Uranus orbit crossing\footnote{\url{http://pluto.jhuapl.edu/mission/whereis\_nh.php}
(Last accessed on 2011 Sep 2)},
and assumed to be constant, although it decreases slightly with the increasing
heliocentric distance.
These simplifications introduce a velocity error not exceeding (1--2) $\kmpers$ 
and thus a 
relative flux error of $\Delta F/F \la 10\%$.
This accuracy is sufficient, given the error bars of the
SDC measurements (see Fig.~\ref{fig:NH_flux_radial}).
The trajectory of New Horizons was taken from the HORIZONS Web-Interface%
\footnote{\url{http://ssd.jpl.nasa.gov/horizons.cgi} (Last accessed on 2013 Sep 9)}.

\subsection{Results}

\begin{figure*}
  \begin{center}
  \includegraphics[width=1.0\textwidth]{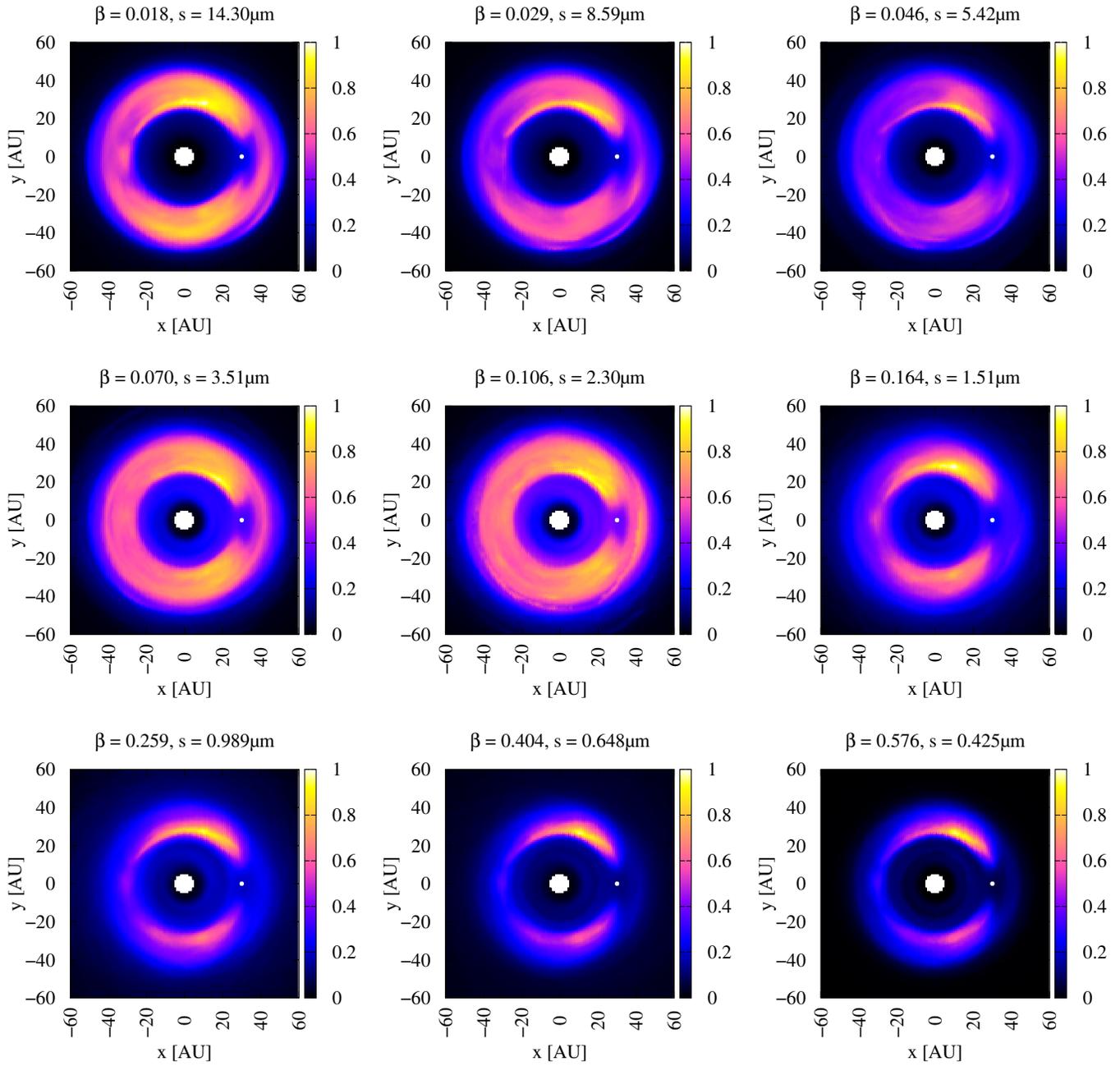}\\
  \end{center}
  \caption{
    Same as Fig.~\ref{fig:map_no_coll} but after applying collisions.
    }
  \label{fig:map}
\end{figure*}

Figure~\ref{fig:map} shows the snapshots produced after applying the collisions.
It demonstrates clearly that resonant structures survive not only for larger, but also 
for smaller grains.
This may seem in contradiction to \citet{Kuchner-Stark-2010} who showed that resonant 
structures will be washed out by destructive collisions,
but they did not inlcude production of grains.
In fact it is the production of grains that is the most important part in our model.
The initial distribution of small grains (bins \#$0$--$4$) gets almost
completely overwritten by small fragments supplied by collisions
of larger grains.

One caveat is that, since we used nine sizes only,
the largest grains (bins \#$5$--$8$) 
do not have parent bodies in our calculation and hence are not replenished.
On the other hand, sufficiently big impactors that could deplete the grains in the upper size 
bins are also absent.
Thus the results presented for the largest grains ($\ga 3\mum$) remain essentially 
collisionless. This explains that the panels in
Figs.~\ref{fig:map_no_coll} and~~\ref{fig:map} for those sizes are nearly identical.
However, the dust impact rates for New Horizons are unaffected, because they are 
dominated by smaller grains.

We now turn to the impact flux on the SDC.
Like the number density, the flux depend on numerous
model parameters describing the populations of EKBOs, collisional cascade physics, and dust 
properties.
Although we can make an educated guess about all of them,
many of these are not really known.
Therefore, it is important to test the robustness of our model 
by varying diverse model parameters.
To this end, we first define our nominal model as follows.
As bulk density we use $\rho = 1.46\gperccm$, the pre-factor of the critical specific 
disruption energy is set to $A = 10^6\ergperg$,
as exponent for the crushing law we adopted the standard value of 
$p = 3.5$.
It is this model that we used in constructing Figs.~\ref{fig:map_no_coll}-\ref{fig:map}.
The fluxes for the nominal model are presented with the thick 
blue line in all panels of Fig.~\ref{fig:NH_flux_radial}.
Then we varied each of the values separately while leaving the others constant.
The individual checks were done as follows:

\begin{figure*}[htb!!]
  \begin{center}
  \includegraphics[width=1.00\textwidth]{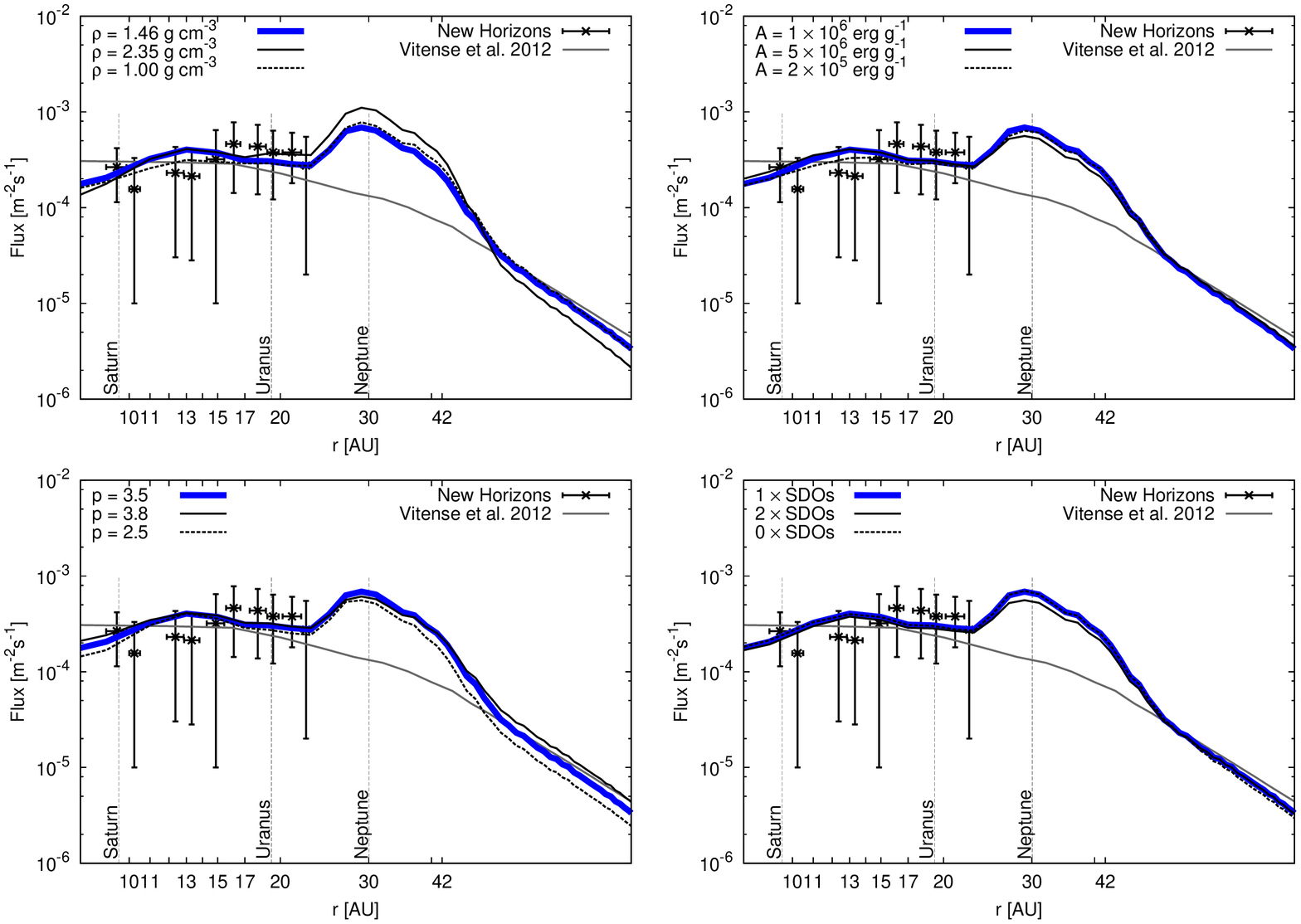}
  \end{center}
  \caption{
    Modeled particle flux (lines)
    fitted to the in-situ measurements of New Horizons (symbols with 1-$\sigma$ error bars). 
    The thick blue line is the same in all four panels and respesents our nominal model
    (ice-silicate mixture with $\rho = 1.46\gperccm$, material strength with
    $A = 10^6\ergperg$ in Eq.~\ref{QD}, the crushing law exponent $p = 3.5$,
    SDO populations from \citeauthor{Vitense-et-al-2010} \citeyear{Vitense-et-al-2010}).
    Individual panels show the dependence of the flux on the
    assumed material composition (top left),
    material strength (top right),
    the crushing law exponent (bottom left),
    and the assumed number of SDOs (bottom right).
    Grey line is a rotationally symmetric model of \citet{Vitense-et-al-2012} for comparison.
    The data are taken from 
    \cite{Poppe-et-al-2010}, \cite{Han-et-al-2011}, and \cite{Szalay-et-al-2013}.
    }
  \label{fig:NH_flux_radial}
\end{figure*}

\begin{enumerate}
\item
 {\em Bulk density and optical properties of dust}.
 As compositions we used the three different material mixtures introduced in 
 Table~\ref{tab:used_betas}.
 The corresponding fluxes are shown in the top left panel of 
 Fig.~\ref{fig:NH_flux_radial}.
 There is a slight trend towards a higher particle flux in the clumps for grains with a lower
 ice content (or with a higher bulk density).
 Yet the difference is much smaller than the data point error bars, and hence the bulk 
 density cannot be constrained further.
 
\item
 {\em Mechanical strength of dust}.
 We have checked the sensitivity of our results to the critical specific 
 energy $Q_\mathrm{D}^*$ which is not reliably known either.
 We tried to replace $A = 10^6\ergperg$ with $A = 5\times 10^6\ergperg$ and $A = 2\times 10^5\ergperg$.
 The results are shown in the top right panel of Fig.~\ref{fig:NH_flux_radial}.
 As expected, increasing the material strength 
 weakens collisional erosion, requiring less material in the outer region to explain the
 observed flux in the Uranus-Saturn region.
 However, the difference between the curves is still minor and there is little
 hope that the New Horizons data could help constraining the material strength~--- 
 unless $A$ deviates from the nominal value by more than an order of magnitude.

\item
 {\em Collisional physics}.
 We investigated the influence of the exponent of the crushing law that determines the size
 distribution of fragments of any destructive collision.
 While assuming $p = 3.5$ in the nominal case, we also tried $p = 2.5$ and $p = 3.8$.
 The bottom left panel shows the result of this variation.
 Here, too, the differences are too small to be measurable by the SDC.

\item
 {\em Parent body populations}.
 One more potentially important uncertainty in our model is related to
 the dust parent bodies. While the true amount and orbital element distributions of classical
 and resonant populations are thought to be known reasonably well, those of 
 the scattered disk objects are not \citep{Vitense-et-al-2010}. Reliable estimates
 for SDOs are not possible because of their high eccentricities and large semimajor axes.
 To check the potential effect, we used the ``true'' SDO population derived by
 \citet{Vitense-et-al-2010} in the nominal model, but also tested the extreme 
 cases of no SDOs and twice as many SDOs as in the nominal model.
 The results are presented in the bottom right panel.
 With twice as many SDOs there is a marginal decrease of dust flux in the clumps.
 Albeit too small to be detectable, the very effect is interesting.
 When dust particles are released from SDOs their semimajor axes and eccentricities are
 higher than if they are released from classical objects. Then, the PR drag decreases both
 $a$ and $e$. Nevertheless, when they finally reach the location of the $3:2$
 resonance, the  eccentricities are still higher than those of the particles released by
 classical objects.
 As a result, the probability of being captured into the mean motion resonance
 is lower \citep{Mustill-Wyatt-2011}, decreasing the dust density in the clumps.
 Moreover, when these highly eccentric particles enter the region between  Saturn and Uranus,
 they have higher mean relative velocity and experience more disruptive collisions,
 which enhances the dust concentration there. However, since we fit the modeled fluxes to the
 SDC measurements at those locations, they are ``nailed'' to the data points, which
 further decreases the height of the flux curve in the clumps.
 
\end{enumerate}

Beside the parameter variation, one more test was needed for the following reason.
The results presented above were obtained by counting the particles with 
all $z$-coordinates,
 implicitly assuming that the dust disk is uniform vertically.
 However, this is not true, and New Horizons has a non-zero inclination 
 of $i_{_\text{NH}} = 2.4^\circ$.
 To check how this could affect the rates, we have done a separate calculation
 where we only considered particles with $0 < z < r \sin i_{_\text{NH}}$.
 After fitting the resulting curve to the available data, we found it to be nealy 
 indistinguishable from the nominal curve.

From all these tests, we argue that
New Horizons should be able to detect the increased impact fluxes when entering the clumps.
The conclusion holds if the poorly known model parameters are varied within physically 
plausible ranges.

\begin{figure}[htb!]
  \begin{center}
  \includegraphics[width=0.49\textwidth]{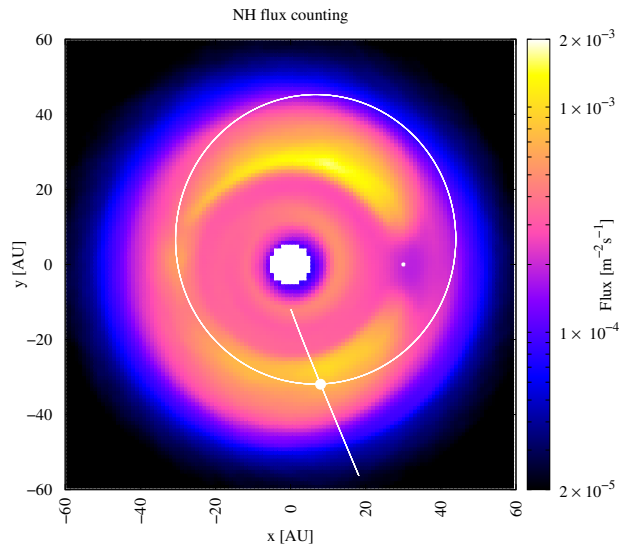}
  \end{center}
  \caption{
    Pole-on view of the modeled particle flux in our nominal model.
    The white ellipse represents the orbit of Pluto, the white line toward bottom left is
    the trajectory of New Horizons, the small white dot is the location of Neptune,
    and the big white dot the intersection between New Horizons and Pluto. 
    }
  \label{fig:NH_flux_xy}
\end{figure}

Figure~\ref{fig:NH_flux_radial} presents the particle flux
averaged over the heliocentric longitude.
A pole-on (and therefore spatially resolved) view is given in Fig.~\ref{fig:NH_flux_xy}.
It reveals that the clumps are not enirely symmetric. 
The dust flux in the Neptune-trailing clump (which will be flown through by New 
Horizons) should be slightly lower than in the leading one.

\section{CONCLUSIONS}\label{sec:conclusion}

In this paper we developed a model to reproduce and predict dust impact rates onto the 
dust detector of New Horizons.  
To this end, we used the debiased EKBO populations from \citet{Vitense-et-al-2010}
and launched $27000$ dust grains of nine sizes.
The particle sizes were chosen in such a way as to cover the entire mass sensitivity range
of the dust detector aboard New Horizons.
We then integrated their orbits by including the gravity of 
the four giant planets, stellar wind and the Poynting-Robertson effect, and recorded the 
position and velocities every orbit of Neptune.
The resulting $2 \times 10^9$ records were used to draw a 
collisionless dust density map.
We then complemented this collisionless model with a post-processing algorithm
to roughly simulate a collisional cascade.

In the collisionless approximation, our model fully reproduces the results
obtained previously with similar methods. The most salient feature of all
collisionless models, pioneered by the \citet{Liou-Zook-1999} study, is a pair of
resonant clumps ahead and behind the Neptune location.
These clumps are dominated by grains that are large enough to be efficiently
captured in resonances with Neptune.

Including collisions has a potential of changing the number density map for small 
(submicrometer-sized) grains dramatically. Since it is these grains that are
predominantly detected by the New Horizons dust detector, the issue is important.
\citet{Kuchner-Stark-2010} made the first attempt to include collisions in
the models of azimuthal structure in the EKB region and showed
that the collisonal elimination of grains should essentially wash out the clumps.
However, they only considered collisional loss of grains.
Our model refines theirs by adding
the collisional replenishment of small grains.
This tends to increase the amount of dust in the clumps almost back to the
level predicted by the collisionless models.

We have calculated the dust impact fluxes onto 
the New Horizons dust detector to find that it should be able to 
detect an increase of particle flux when it enters the region of the clump ahead the position 
of Neptune.
The dust flux should increase by a factor of two or three.
By varying the assumed material composition, mechanical strength of dust
and the exponent of the crushing law, we checked
that this result is rather robust, making our predictions quite certain.
At the same time, this implies that New Horizons will probably not be able to provide
additional contraints on the dust properties.
The same applies to uncertainties stemming from a poorly known distribution of the dust parent 
bodies, especially of the scattered population of the Kuiper belt objects.
Including or excluding this population from the simulations does not affect the predicted 
impact rates considerably.

\section{DISCUSSION}\label{sec:discussion}

Although we did check that our predictions should not strongly depend on the unknown 
properties of the dust grains and detailed distributions of dust-producing Kuiper belt 
objects, the model itself necessarily involves a number of simplifications, 
assumptions, and omissions that may affect the results.
Apart from standard assumptions, such as that of spherical particles or using the 
Mie theory for radiation pressure calculations,
caveats exist in our simulations of the collisional cascade.
While we improved previous EKB dust models, complementing
collisional depletion by collisional production,
we assumed, for instance, equal-size impactors.
Collisions between different-sized impactors would lead 
to somewhat higher collisional velocities and rates an thus increase the total 
number of collisionally produced small particles, 
resulting in a higher dust flux.
Nevertheless,  since we are fitting our model to the data of New Horizons,
the ``missing'' flux will automatically be compensated by a larger 
$\xi$ in Eq.~\ref{eq:xi}.

Even more importantly,
we did not include radiation pressure on newly produced particles.
After a disruptive collision between large grains, small particles gain orbital energy
from radiation pressure which sends them onto a different, usually eccentric and 
wide, orbit. 
This may spread the clumps, decreasing the dust density enhancement compared to what 
our model predicts, closer to what previous models without collisional production suggest.
Nevertheless, we do not think
that the structures would be completely washed away.  First, after
spreading, the particles will drift back inward and many will be re-captured
into resonances.  Even for small particles (sizes \#2 and \#3) a slight
enhancement can be seen at the location of resonances.  Second, the highest
collision rates are in the clumps, hence there is an enhanced 
production and thus a higher number density of fine grains there \citep{Wyatt-2006}.
Third, the pericenters of collisionally produced dust  
particles will always lie at their birth point and thus~--- in contrast to
their apocenters~--- are not subject to radation pressure spreading
(see discussion at the end of Sect.~2.2).
Altogether, we believe the ``true'' distribution of dust is
somewhere between previous models and ours, 
perhaps closer to the latter.
This is to say that New Horizons might measure a more 
moderate enhancement of the impact rate when traversing the clump than our model predicts.

There are also other factors that might decrease the enhancement.
One of them is the ``sporadic'' dust background stemming from sources other than EKBOs,
e.g. Trojans, Centaurs, or comets of Schwassmann-Wachmann type \citep{Landgraf-et-al-2002}.
Some contribution can also be made by interstellar grains,
especially given that New Horizons is facing the upstream direction of the 
interstellar dust flow.
However, \citet{Altobelli-et-al-2007} showed that the interstellar grains of size $s\approx 
0.4\mum$
contribute to the overall dust flux by $F = 2\times 10^{-5}\persmpers$ which is 
an order of magnitude below the measurements of the SDC and therefore can be neglected.
The possibilities to make the clumps {\em more pronounced} than predicted 
here are
more limited. Should the enhancement seen by the SDC be stronger than expected, this could be 
attributed, for instance, to a lower critical disruption energy or to a higher bulk density
of dust particles.

\begin{acknowledgements}
We would like to thank Martin Reidemeister for a helpful discussion of several numerical aspects of this work.
Useful comments by the anonymous reviewer are appreciated.
This research was partly supported by the
\emph{Deut\-sche For\-schungs\-ge\-mein\-schaft} (DFG), projects number Lo~1715/1-1
and Kr~2164/10-1.
\end{acknowledgements}


\newcommand{\AAp}      {Astron. Astrophys.}
\newcommand{\AJ}       {Astron. J.}
\newcommand{\AO}       {App. Optics}
\newcommand{\ApJ}      {Astrophys. J.}
\newcommand{\EPS}      {Earth, Planets and Space}

\end{document}